\documentclass[superscriptaddress,aps,prl,floats,twocolumn,twoside,floatfix,show
pacs]{revtex4}
\usepackage{epsfig}
\usepackage{amsmath}
\usepackage{amssymb}

\begin{document}

\title{Statistical mechanics of combinatorial auctions}

\author{Tobias Galla}
\affiliation{The Abdus Salam International Centre for Theoretical Physics, 
Strada Costiera 11, 34014 Trieste, Italy}
\affiliation{INFM/CNR SISSA Unit, Via Beirut 2-4, 34014 Trieste, Italy}
\author{Michele Leone}
\affiliation{Institute for Scientific Interchange, Viale
  S. Severo 65, 10133 Torino, Italy}
\author{Matteo Marsili}
\affiliation{The Abdus Salam International Centre for Theoretical Physics, 
Strada Costiera 11, 34014 Trieste, Italy}
\author{Mauro Sellitto}
\affiliation{The Abdus Salam International Centre for Theoretical Physics, 
Strada Costiera 11, 34014 Trieste, Italy}
\affiliation{Institute for Scientific Interchange, Viale
  S. Severo 65, 10133 Torino, Italy}
\author{Martin Weigt}
\affiliation{Institute for Scientific Interchange, Viale
  S. Severo 65, 10133 Torino, Italy}
\author{Riccardo Zecchina}
\affiliation{The Abdus Salam International Centre for Theoretical Physics, 
Strada Costiera 11, 34014 Trieste, Italy}
\date{\today}

\begin{abstract}
  Combinatorial auctions are formulated as frustrated lattice gases on
  sparse random graphs, allowing the determination of the optimal
  revenue by methods of statistical physics.  Transitions between
  computationally easy and hard regimes are found and interpreted in
  terms of the geometric structure of the space of solutions. We
  introduce an iterative algorithm to solve intermediate and large
  instances, and discuss competing states of optimal revenue and
  maximal number of satisfied bidders. The algorithm can be
  generalized to the hard phase and to more sophisticated auction
  protocols.
\end{abstract}

\maketitle

Auctions constitute an important part of economic activity
\cite{Krishna}.  With today's pronounced role of e-commerce and the
use of the internet as a world-wide market place, fundamental changes
have occurred in the use of auctions. Their popularity has increased
due to freely accessible web services, and both the range and the
nature of objects which can be bought and sold have diversified.
Single-item auction protocols are however inappropriate when the
number of objects to be sold is large, especially if buyers are
interested in bundles of objects with complementary features. In such
cases, it is preferrable to allow buyers to bid on {\em combinations}
of objects instead of single items.  Such auctions are referred to as
{\em combinatorial auctions} (CA).  Initially motivated by the problem
of airport time slot allocation and by the distribution of radio
spectrum licenses, CA are now widely used in a variety of
contexts~\cite{MIT}.  Finding the maximum auctioneer's payoff
allocation to a given CA is a computationally hard problem requiring
exponential time resources.  Fast and efficient heuristic algorithms
are thus needed to identify optimal or close-to-optimal allocations,
and models of CA are here of interest as theoretical benchmarks
\cite{Sandholm,vohra}.

In this Letter, we will first map a generic CA problem onto a
geometrically constrained lattice gas defined on a sparse random
graph, coupled to a local chemical potential. This allows us to relate
the computational complexity of determining optimal allocations to the
glassy behaviour induced by the geometric frustration of the system.
The latter here results from conflicting bids.  The statistical
mechanics approach to Bethe lattice glass models \cite{rivoire} then
allows us to characterize the typical properties of large random
instances of CA, and to formulate algorithms for solving single
instances.

A simple CA model describes $N$ players and a total set of $M$ items
to be sold. Each player $i$ bids for an individual subset
$A_i\subseteq \{1,\dots,M\}$ for which he is willing to pay a price
$\nu_i>0$. It is understood here that the player is only interested in
the full set $A_i$, but not in any proper subset.  The bid of player
$i$ is thus given by the pair $(A_i,\nu_i)$. More complex CA where
bidders submit lists of $n_i\ge 1$ subset-price pairs,
$\{(A_i^\ell,\nu_i^\ell)\}_{l=1,...,n_i}$ nested by logical ORs or
XORs can be reduced to the case of single-item bids, with a possibly
enlarged set of bidders and items \cite{Nisan}.  Hence we shall
confine our discussion to the $n_i=1$ case of single-subset bids.

Let us indicate a successful bid by $x_i=1$, and an unsuccessful one
by $x_i=0$.  Then the {\em winner determination problem} (WDP)
consists in finding a `configuration' $\mathbf{x}=(x_1,\dots,x_N)\in
\{0,1\}^N$, which maximizes the auctioneer's {\it revenue} $R=\sum_i
\nu_i x_i$ and respects the constraint that each item be sold at most
once, i.e., that $x_ix_j=0$ whenever $A_i\cap A_j\neq 0$ for a pair
$i\neq j$.  It is also interesting to discuss how far the optimal
solution is from a configuration maximizing the number $N_s = \sum_i
x_i$ of satisfied bidders. These two tasks are usually conflicting,
and can be formulated as a {\it multi-objective optimization} problem
whose outcome depends on the optimization priority. In particular we
will compare results of procedures first maximizing $R$ and then
$N_s$, and vice versa.

Deciding which bids $(A_i,\nu_i)$ are successful ($x_i=1$) is a
non-trivial optimization problem, because bids can overlap. If
$A_i\cap A_j\neq\emptyset$ for some $i\neq j$, the two bidders $i$ and
$j$ cannot simultaneously be successful.  Even an agent with a high
bid $\nu_i$ is not guaranteed to win, as it may be advantageous for
the auctioneer to allocate the items contained in $A_{i}$ to a number
of other agents having a higher aggregate revenue. Indeed the CA
optimal allocation problem is NP-complete, as can be demonstrated by
its equivalence to the weighted set packing problem \cite{PAP}.
NP-completeness, however, refers to a worst-case scenario and does not
necessarily imply that one cannot find the optimal solution for some
large real-life CA problems.
\begin{figure}
\begin{center}
\includegraphics[width=4.5cm]{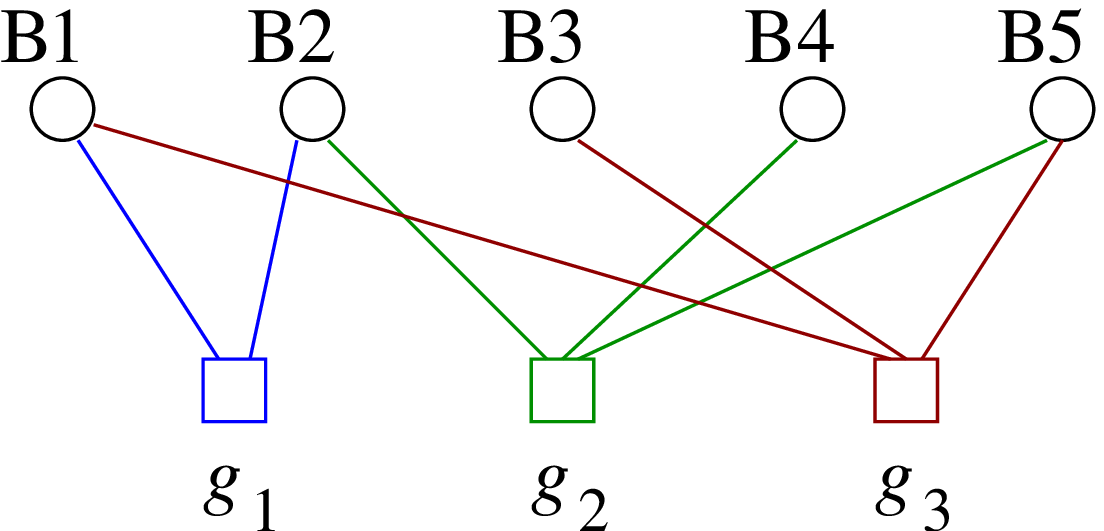}
\includegraphics[width=3cm]{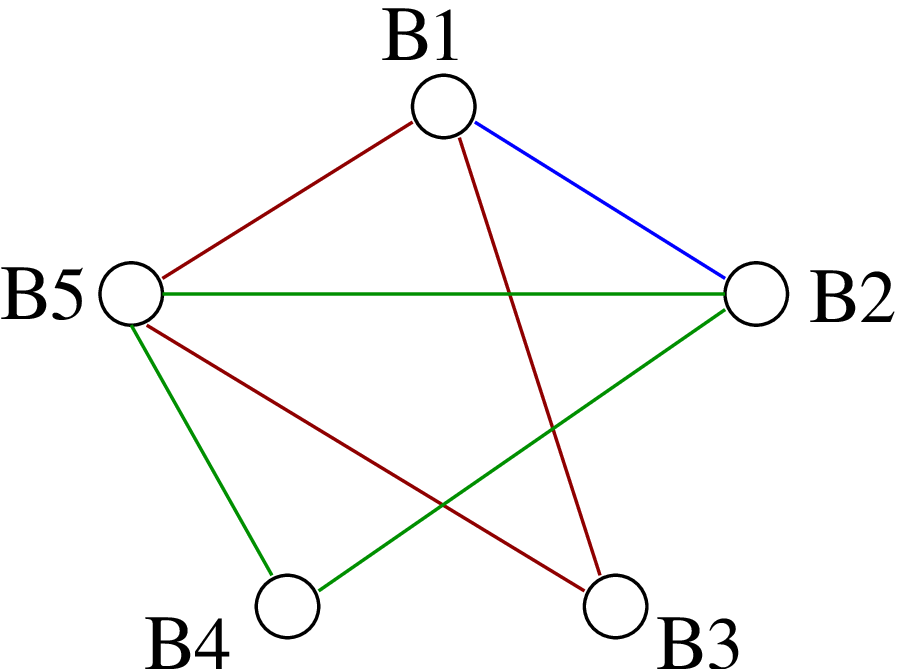}
\caption{(Color online) Example of a CA 
        interaction network: {\sl factor graph} (left),
        and corresponding {\sl conflict graph}
        (right). Bidders $B_i$ are indicated as circles, items $g_a$
        as squares attached via links to all the bidders wanting to
        pay for them. In the conflict graph bidders are
        connected by a link if and only if they share at least one
        item. Any item of degree $k$ in the factor graph
        corresponds to a $k$-clique of vertices in the conflict graph.
        $I_a$ is the set of players bidding for item $a$.
        \label{fig:prima}}
\end{center}
\end{figure}
Here, we focus on a simple probabilistic model where each player
submits a single bid. This setup retains the same level of
computational complexity as more general cases, and the WDP can be
reduced to computing the ground state of a lattice gas defined
on a suitable random graph, where particles residing on vertices are
subject to a geometric constraint as well as to a local chemical
potential.

{\bf The Model ---} Let us take $N$ players and $M=\alpha N$ items,
with $\alpha={\cal O}(1)$.  Player $i$ bids for a subset $A_i$
containing $\ell_i$ items, and any item $a$ in turn is chosen by
$k_a$ agents. Any realization of this bidding structure can then be
represented by a {\em factor graph} containing bidder and item nodes,
cf. Fig.~\ref{fig:prima}. A compact representation is given by
the {\sl conflict graph} (CG), in which two players $i$ and $j$ are
linked whenever $A_i\cap A_j\neq\emptyset$.  More complex bidding
protocols can be cast in the same formalism \cite{in-prep}.  

The CA can be mapped to a {\sl lattice gas} of particles on the CG,
with occupation variables $x_i\in\{0,1\}$ representing the bidders'
success/failure. The Hamiltonian ${\cal H}=-R=-\sum_i \nu_i x_i$
describes a coupling of $x_i$ to a local field $\nu_i$, and it can be
associated with a formal inverse temperature $\beta$. A chemical
potential $\mu$ allows to control the particle number $N_s=\sum_i
x_i$. The grand-canonical partition function of such a lattice-gas
reads
\begin{eqnarray}
  Z = \sum_{\scriptscriptstyle \mathbf{x}\in \{0,1\}^N}\!\! \exp
  \left(\sum_{i=1}^N (\mu + \beta \nu_i) x_i \right)
  \prod_{(i,j)\in C} (1-{x_i x_j}) ,
  \label{Z}
\end{eqnarray}
where $C$ is the set of edges of the CG. The product over $C$
implements the compatibility constraint that each item can be sold at
most once, and can be interpreted as a volume exclusion for
neighbouring lattice sites. Temperature and chemical potential can be
used to select compatible solutions in configuration space: {\sl (i)}
the choice $\beta \to \infty$ and $\mu = 0$ corresponds to
configurations maximizing $R$ independently of $N_s$, {\sl (ii)}
$\beta = 0$ and $\mu \to \infty$ selects configurations maximizing
$N_s$ only, {\sl (iii)} $\beta \to \infty$ {\sl before} $\mu \to
\infty$: maximizes $N_s$ within the subset of configurations of
maximal revenue, {\sl (iv)} $\mu \to \infty$ {\sl before} $\beta \to
\infty$: maximizes $R$ within the subset of compatible configurations
of maximum number of satisfied bidders. The maximal revenue can thus
be computed as $R^* = \lim_{\beta \to\infty} \partial_{\beta} \ln Z$
at finite $\mu$, and the maximum $N_s^* = \lim_{\mu \to \infty}
\partial_{\mu} \ln Z$ at finite $\beta$.

\begin{figure}
\begin{center}
  {\includegraphics[width=5cm]{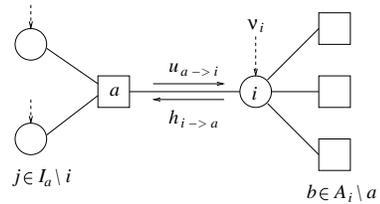}}
\caption{Illustration of cavity biases and fields.}
\label{fig:wp}
\end{center}
\end{figure}

{\bf The cavity approach --- } The search for optimal configurations
can be performed efficiently via algorithms based on the {\em cavity
  method}. For a general presentation of the method see e.g.
\cite{MZ,book} and references therein. Following \cite{book}, one can
introduce cavity biases $\{u_{a\to i}\}$ and cavity fields $\{h_{j\to
  a}\}$ associated with the links of the factor graph of a single CA
instance as illustrated in Fig.~\ref{fig:wp}. The cavity bias $u_{a\to
  i}$ measures the likelihood that item $a$ is already assigned to
another bidder $k \in I_a\setminus i$, and cannot be sold to $i$. The
cavity field $h_{i\to a}$ is then given by the re-weighted price minus
the sum of all $u_{b \to i}$ arriving from $b\in A_i$ except from $a$.
It measures the likelihood that the bidder would win if his bid did
not contain item $a$. The resulting self-consistent equations are the
basis of the so-called {\em belief propagation} (BP) algorithm:
\begin{eqnarray}
  u_{a\to i} 
  &=& \hat u(\{h_{k \to a}\})
  = \frac{1}{\beta + \mu} \ln \left( 1 + \sum_{k \in
  I_a\setminus i} e^{(\beta + \mu) h_{k \to a}} \right) \nonumber \\
  h_{i\to a} 
  &=& \hat h( \nu_i, \{u_{b\to i}\} )
  = \frac{\mu + \beta \nu_i}{\beta + \mu} - \sum_{b\in
  A_i \setminus a} u_{b\to i}.
  \label{BP}
\end{eqnarray}
If the iteration of Eqs.~(\ref{BP}) converges to a fixed point, we can
estimate the probability that bid $i$ becomes satisfied: 
\begin{equation}
P_{i} \equiv P(x_i=1) \simeq 
\frac{ \exp\{(\mu+\beta)H_i\}}{1 +  \exp\{(\mu+\beta)H_i\}}
\end{equation}
with an effective local field $H_i = (\mu + \beta \nu_i)/(\beta +
\mu) - \sum_{b\in A_i} u_{b\to i}$. The payoff estimate becomes $\sum_i R =
\nu_i P_{i}$, and the expected number of satisfied bids is $N_{s} =
\sum_i P_{i}$.

The maximization of these quantities can be achieved by
tuning $\beta$ and $\mu$ as described after Eq.~(\ref{Z}).  In
principle, the limit $\beta\to\infty$ at $\mu=0$ -- corresponding to a
maximum revenue -- can be taken directly in Eqs.~(\ref{BP}), and has a
simple interpretation. In this case $u_{a \to i}$ is a {\em warning}
to bidder $i$ that he needs to allocate at least a price $u_{a \to i}$
to item $a$ in order to outbid the other players in $I_a$.
Consequently, in order to secure his subset $A_i$, he has to offer at
least $\sum_{a\in A_i} u_{a \to i}$. The field $H_i$ is thus given by
the difference between the amount $i$ is willing to pay and the
minimal amount required for $i$ to win. If $H_i>0$ the bidder
will be successful ($x_i=1$) with probability $1$ (i.e. in all such
solutions), whereas $H_i<0$ indicates a loosing bid. Players with
$H_i=0$ are successful in some optimal assignments, and unsuccessful
in others. In order to deal with the latter degeneracy, one has to
resort to a more subtle limit of Eqs.~(\ref{BP}), allowing for fields
vanishing proportionally to the formal temperature.

Eqs.~(\ref{BP}) are valid for any fixed choice of the graphs, prices
and bids, i.e. the effective fields can be determined for {\it any
  specific CA instance}. They can be solved iteratively in ${\cal
  O}(N)$ steps whenever the iteration converges. The fields $\{H_i\}$ at
convergence of the BP iteration can be used as input for an iterative
search/decimation procedure, where items are assigned iteratively to
bidders with highest $H_i$, and BP equations are then iterated for the
reduced problem containing only variables not yet fixed in the
previous step. This procedure runs in ${\cal O}(N \log N)$ steps if a finite
fraction of items is assigned at each step.

{\bf Solution for typical cases ---} To extract analytical estimates
for the CA outcomes we average Eqs.~(\ref{BP}) over given random
factor-graph and price ensembles. In doing so we obtain
self-consistent integral equations for the order parameters, i.e. the
histograms of both the cavity biases ($Q(u)$) and the cavity fields
($P(h)$). In the so-called replica symmetric (RS) phase
\cite{MZ,book}, these equations are
\begin{eqnarray}
  P(h)&=&\sum_{\ell=0}^\infty \tilde p_\ell \int \prod_{t=1}^{\ell} du_t
  Q(u_t) \int d\nu P(\nu|\ell+1)\delta(h-{\hat h}) \nonumber \\
  Q(u)&=&\sum_{k=1}^\infty \tilde q_k \int \prod_{t=1}^k dh_t P(h_t)
  \delta(u-{\hat u})
\label{pop}
\end{eqnarray}
where $P(\nu|\ell+1)$ is the distribution of prices $\nu$ for agents
bidding for sets of $\ell+1$ objects, ${\hat h}$, ${\hat u}$ are given
by (\ref{BP}), $\tilde p_\ell = (\ell+1)
p_{\ell+1}/\langle\ell\rangle$ and $\tilde q_k = (k+1) q_{k+1}/\langle
k\rangle$, where $p_\ell$ and $q_k$ are the degree distributions of
the bidders and items in the factor graph.  Eqs.~(\ref{pop}) are
solved via population dynamics \cite{book}. Here we focus on two
simple non-trivial CA ensembles. In both cases, every player selects
each item independently with probability $z/M$. The probability that a
player wants $\ell$ items is Poissonian, $p_\ell={\rm e}^{-z}
z^{\ell}/\ell \, !$, for large $M$. Similarly, the probability that a
given item is chosen by $k$ players is $ q_k = {\rm e}^{-\lambda}
\lambda^k/k \,!$ for large $N$, where $\lambda=z/\alpha$. The
considered ensembles differ in the prices the bidders are willing to
pay:

\begin{figure}[t!]
\begin{center}
  \includegraphics[width=8cm]{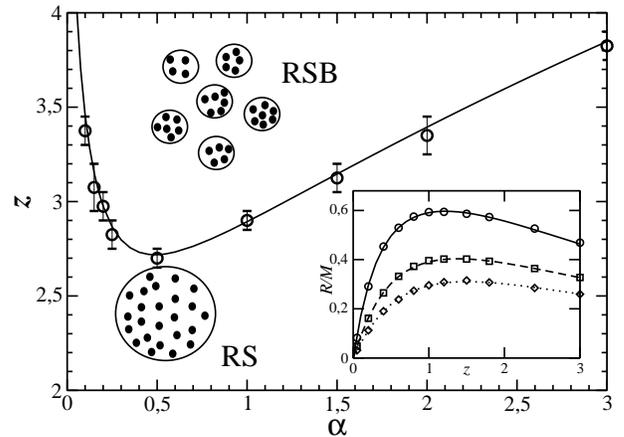}
\caption{Phase diagram for constant-price CAs in the
  $(\alpha,z)$-plane: in the lower region, all solutions
  (maximum-revenue configurations) belong to a single cluster, and RS
  equations are believed to give the exact revenue result. In the
  upper region the RS solution is not correct anymore.  The symbols
  mark the point where BP stops to converge on $\sim 3\%$ of the spins
  ($N=5000$, averaged over $20$ instances), and are consistent with
  the line at which RS breaks. {\it Inset:} Maximal
  revenue per item vs. $z$ for $\alpha=0.5,1,1.5$ from top to bottom.
  Average RS results (lines) are compared to simulated-annealing data
  on single CA instances (markers). Results obtained by iterating BP
  on the same CA instances (not shown) are consistent with these
  curves.}
\label{fig:constant_price}
\end{center}
\end{figure}

{\em (i) Constant prices ---} Every bidder who bids for at least one
object offers the same price $\nu_i\equiv 1$. BP-results from Eqs.
(\ref{pop}) are shown in the inset of Fig.~\ref{fig:constant_price}
together with optimal revenue estimates from simulated annealing.
Note the maximum of the revenue for intermediate $z$. At small $z$
many items are not element of any bid and do not contribute to the
revenue, while at large $z$ items are desired by multiple players
resulting in conflicts which restrict the revenue. The main body of
the figure shows analytical results \cite{in-prep} on the validity of
RS. Above the phase boundary conflicts induce replica-symmetry
breaking (RSB), i.e. a clustering of CA solutions into disconnected
sets. RS results cannot be trusted beyond that point and multiple
metastable states are found. Note that RSB in the typical case
corresponds to non-convergence of BP on relatively large, randomly
generated single instances of CAs. This is confirmed by the symbols,
which mark the points where BP stops to converge.

\begin{figure}[t!]
\begin{center}
  \includegraphics[width=8cm]{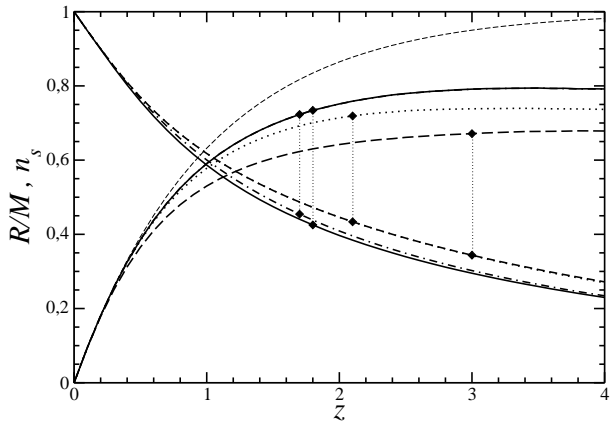}
\caption{Revenue (increasing curves) and fraction $n_s$ of satisfied players
  in the set of players with non-emtpy bids (decreasing curves) for
  the four optimization problems ($\alpha=1$). Maximisation order is
  as follows: {\it (i)} $R$ only (full lines), {\it (ii)} $N_s$ only
  (dashed lines), {\it (iii)} $R$ before $N_s$ (dash-dotted lines),
  {\it (iv)} $N_s$ before $R$ (dotted lines). Revenues for (i) and
  (iii) indistiguishable, $n_s$ for (ii) and (iv)
  indistinguishable. Symbols mark the points where RS breaks down, RS
  results are approximations at large $z$. The upper most increasing
  line marks the fraction of items which are part of at least one bid,
  i.e. the theoretical maximum number of auctionable objects.}
\label{linear}
\end{center}
\end{figure}

{\em (ii) Linear prices ---} In this second, potentially more
realistic case the price of a bid is assumed to equal the size of the
desired subset, i.e. the price per item is constant. Qualitatively,
the revenue curves show a behaviour similar to the constant-price
case, cf.~Fig.~\ref{linear}. For small $z$, the revenue is very close
to $1-e^{-\lambda}$, the fraction of items wanted by at least one
bidder. Even if there is little frustration, it is not possible to
sell all items as it would be optimal for the auctioneer in this case
\cite{note2}. For larger $z$, replica symmetry breaking sets in as a
result of increasing frustration, and asymptotically the revenue
curves are expected to decrease again with $z$. The increasing level
of frustration can also be seen in the fraction of successful players
in the group of players placing non-empty bids. As also shown in
Fig.~\ref{linear}, this fraction is found to decrease monotonously
with $z$, i.e. with the average number of items wanted by a bidder (at
fixed $\alpha$).

{\em (iii) Other ensembles---} Other CA ensembles give rise to even
more complex phase diagrams. For instance, in the elementary case of
each player wanting a fixed number $K$ of randomly chosen items, all
at the same price (corresponding to $\nu_i=const$), one finds that the
space of optimal solutions is divided into an exponential number of
geometrically distant clusters even for values of $\alpha$ at which
all items can be assigned without frustration. This corresponds to a
1-RSB mechanism observed in other hard combinatorial problems
\cite{book}.  In such a region (e.g. $\alpha \in [.56, .59]$ for
$K=5$) message-passing algorithms~\cite{MZ,in-prep} provide efficient
techniques for finding optimal assignments.

The extension to CAs with generic price distribution is
straightforward and will be presented elsewhere~\cite{in-prep}.

Let us finally mention that the case of linear prices allows one to
study multi-objective strategies for solving the WDP going beyond
standard CA theory which is mostly concerned with revenue
maximization. As discussed before, one could also aim to maximize the
number of satisfied players. As shown in Fig.~\ref{linear}, doing this
after maximizing the revenue has only very little effect on $N_s$. On
the other hand, maximizing $N_s$ independently of the revenue leads to
a substantial decrease in $R$.  The latter can be partially cured by
maximizing the revenue within the set of winner assignments of maximal
$N_s$, but also there the revenue remains well below its maximum. A
suitable strategy of taking into account both $R$ and $N_s$ might be
to send both $\beta$ and $\mu$ to $\infty$, with a fixed ratio
$\beta/\mu$ controling the relative importance of either optimization
task.  By varying it one can continuously tune the results between the
extremes given before.

To conclude, we have presented a statistical physics approach to the
CA problem and introduced an iterative algorithm for solving the WDP
of large instances of simple CAs.  A relatively large body of
literature has been devoted to algorithms for solving WDP, such as
branch-and-bound and mixed integer programming techniques
(see~\cite{Sandholm,vohra} and references therein).  It would be
interesting to compare the performance of message-passing algorithms
with the existing approaches to WDP, as it has been done recently for
satisfiability problems in Refs.~\cite{erik,mikko}.

This work was supported by the {\small EC} Human Potential Programme,
contracts {\small HPRN-CT-2002-00319}, {\small STIPCO}, {\small EU
NEST No. 516446}, {\small COMPLEXMARKETS} and {\small EU IP} No. 1935,
{\small EVERGROW}.

\end{document}